\documentclass[aps,twocolumn,superscriptaddress,prl,fleqn,showpacs,nofootinbib,preprintnumbers]{revtex4-1}
\usepackage{amssymb,amsmath,epsfig}

\renewcommand{\d}{\mathrm{d}}
\renewcommand{\bm}[1]{\mbox{\boldmath $#1$}}

\newcommand{\mc}[1]{\mathcal{#1}}
\newcommand{\Tr}{\text{Tr}}

\renewcommand{\d}{\mathrm{d}}
\newcommand{\sT}{{\scriptscriptstyle T}}
\newcommand{\pT}{\bm{p}_\sT}
\newcommand{\kT}{\bm{k}_\sT}
\newcommand{\qT}{\bm{q}_\sT}

\renewcommand{\Re}{\text{Re}}

\begin{document}

\title{Determining the Higgs spin and parity in the diphoton decay channel}

\author{Dani\"el Boer}
%\email{D.Boer@rug.nl}
\affiliation{Theory Group, KVI, University of Groningen,
Zernikelaan 25, NL-9747 AA Groningen, The Netherlands}

\author{Wilco J. den Dunnen}
%\email{wilco.den-dunnen@uni-tuebingen.de}
\affiliation{Institute for Theoretical Physics,
                Universit\"{a}t T\"{u}bingen,
                Auf der Morgenstelle 14,
                D-72076 T\"{u}bingen, Germany}
                
\author{Cristian Pisano}
%\email{cristian.pisano@ca.infn.it}
\affiliation{Dipartimento di Fisica, 
Universit\`a di Cagliari, 
and INFN, Sezione di Cagliari, 
I-09042 Monserrato (CA), Italy}

\author{Marc Schlegel}
%\email{marc.schlegel@uni-tuebingen.de}
\affiliation{Institute for Theoretical Physics,
                Universit\"{a}t T\"{u}bingen,
                Auf der Morgenstelle 14,
                D-72076 T\"{u}bingen, Germany}

\begin{abstract}
We calculate the diphoton distribution in the decay of arbitrary
spin-0 and spin-2 bosons produced from gluon fusion, taking into account the fact that
gluons inside an unpolarized proton are generally linearly polarized.
The gluon polarization brings about a difference in the transverse momentum distribution 
of positive and negative parity states. 
At the same time, it causes the azimuthal distribution of the photon pair to be 
non-isotropic for several spin-2 coupling hypotheses, 
allowing one to distinguish these from the isotropic scalar and pseudoscalar distributions.

\end{abstract}

\pacs{12.38.-t; 13.85.Ni; 13.88.+e}
\date{\today}

\maketitle

Last year July it was announced that a new boson  
with a mass around 125-126 GeV was observed by both the ATLAS \cite{:2012gk} and CMS \cite{:2012gu} collaborations.
An excess of events was observed in $\gamma\gamma$, $ZZ^*$ and $WW^*$ production 
from proton-proton collisions at a center of mass energy of 7 and 8 TeV.
The observed excess is consistent, within uncertainties, with the production and decay of
the Standard Model (SM) Higgs boson.

Now that the existence of a new particle has been established, both collaborations have begun the
determination of its spin and parity. 
Both ATLAS \cite{ATLAS-gammagammaupdate,ATLAS-ZZupdate,ATLAS-WWupdate} and 
CMS \cite{CMS-ZZupdate,CMS-WWupdate} set approximately 3 $\sigma$ exclusions on 
the $J^P=0^-$ scenario using the $ZZ^*$ channel and the $2_m^+$ hypothesis\footnote{
The coupling of a spin-2 boson to gauge bosons can be realized in multiple ways.
We will use the standard notation in which $2_m^+$ denotes a spin-2 boson with
minimal (lowest dimensional) coupling, which is uniquely defined.}
starts to be disfavored at the 1-3 $\sigma$ level in the $\gamma\gamma$, $ZZ^*$ and 
$WW^*$ channels.
As the decay of a pure spin-1 state to two photons is not allowed according to the Landau-Yang theorem 
\cite{Landau:1948kw,Yang:1950rg}, the $\gamma\gamma$ channel is being used to distinguish
between spin-0 and spin-2 only.
In the $ZZ^*$ and $WW^*$ decay channels, the spin-1 option should also be considered. 

Even though the number of events is much larger in the $\gamma\gamma$ channel, the ability to 
distinguish spin-0 from spin-2 is \emph{not} much better than in the $ZZ^*$ channel.
The reason is that in the $\gamma\gamma$ channel only the distribution of the
polar angle $\theta$ is considered \cite{ATLAS-gammagammaupdate, Ellis:2012jv}.
The spin-0 and spin-2 hypotheses are not very different in this variable after experimental acceptance 
cuts \cite{ATLAS-gammagammaupdate}, leading to a small discriminating power.
The determination of the parity using only this angle is even impossible,
as the distributions of $0^+$ and $0^-$ are exactly equal
and the same holds true for the $2_h^\pm$ scenarios\footnote{
The $h$ subscript indicates that the spin-2 boson couples through a higher-dimensional coupling.
There are multiple higher-dimensional couplings possible.
We follow the convention of Ref.\ \cite{Gao:2010qx,Bolognesi:2012mm} for $2_h^\pm$.}.

In this letter we demonstrate that one can also differentiate between the different
spin scenarios in the $\gamma\gamma$ channel, 
by studying the dependence on the \emph{azimuthal} angle $\phi$ in the Collins-Soper frame \cite{Collins:1977iv},
which is the diphoton restframe with the $\hat{x}\hat{z}$-plane spanned by the 3-momenta of the colliding
protons and the $\hat{x}$-axis set by their bisector. 
Moreover, different spin-2 coupling hypotheses that have an equal $\theta$ dependence 
can be distinguished from each other using the $\phi$ distribution, 
enhancing the analyzing potential of this channel.
Apart from that, we update predictions for the transverse momentum distribution 
\cite{Boer:2011kf,denDunnen:2012ym} which can, in principle, be used to distinguish the different parity 
states $0^+$ from $0^-$ and $2_h^+$ from $2_h^-$ in the $\gamma\gamma$ channel.
Azimuthal angular distributions have been discussed for spin-0 and spin-2 ``Higgs''
production from vector-boson fusion \cite{Hagiwara:2009wt,Andersen:2010zx,Frank:2012wh}, 
but not yet from gluon fusion and not including linear polarization.

A non-trivial $\phi$ distribution in the decay of spin-2 bosons produced from gluon fusion
can be caused by the fact that gluons in an unpolarized proton are generally linearly polarized.
The degree of gluon polarization can be calculated using perturbative QCD (pQCD) for
transverse momentum of the gluon much larger than the proton mass and is found to be large.
For small transverse momentum pQCD cannot be used to calculate the degree of polarization, but 
this lack of knowledge turns out to be of little influence on the final $\phi$ distribution,
which is mostly dominated by the perturbative part. 

The effects of gluon polarization can be described in the framework of 
Transverse Momentum Dependent (TMD) factorization.
In that framework, the full $pp\to \gamma\gamma X$ cross section is split into a partonic
$gg\to \gamma\gamma$ cross section and two TMD gluon correlators, which describe the 
distribution of gluons inside a proton as a function of not only its momentum along the
direction of the proton, but also transverse to it.
More specifically, the differential cross section for the inclusive 
production of a photon pair from gluon-gluon fusion is written as \cite{Sun:2011iw,Ma:2012hh},
\begin{multline}\label{eq:factformula}
\frac{\d\sigma}{\d^4 q \d \Omega}
  \propto 
  \int\!\! \d^{2}\pT \d^{2}\kT
  \delta^{2}(\pT + \kT - \qT)
  \mc{M}_{\mu\rho \kappa\lambda}
  \left(\mc{M}_{\nu\sigma}^{\quad\kappa\lambda}\right)^*
  \\
  \Phi_g^{\mu\nu}(x_1,\pT,\zeta_1,\mu)\,
  \Phi_g^{\rho\sigma}(x_2,\kT,\zeta_2,\mu),
\end{multline}
with the longitudinal momentum fractions $x_1={q\cdot P_2}/{P_1\cdot P_2}$
and $x_2={q\cdot P_1}/{P_1\cdot P_2}$, $q$ the momentum of the photon pair, 
$\mc{M}$ the $gg\to \gamma\gamma$ partonic hard scattering matrix element
and $\Phi$ the following unpolarized proton gluon TMD correlator,
\begin{multline}\label{eq:TMDcorrelator}
\Phi_g^{\mu\nu}(x,\pT,\zeta,\mu) \equiv
	      2\int \frac{\d(\xi\cdot P)\, \d^2 \xi_\sT}{(x P\cdot n)^2 (2\pi)^3}
	      e^{i ( xP + p_\sT) \cdot \xi}\\
	\qquad \qquad \quad \Tr_c \Big[ \langle P| F^{n\nu}(0)\,
	      \mc{U}_{[0,\xi]}^{n[\text{--}]}\, F^{n\mu}(\xi)\, \mc{U}_{[\xi,0]}^{n[\text{--}]}
	      |P\rangle \Big]_{\xi \cdot P^\prime = 0}\\
=	-\frac{1}{2x} \bigg \{g_\sT^{\mu\nu} f_1^g
	-\bigg(\frac{p_\sT^\mu p_\sT^\nu}{M_p^2}\,
	{+}\,g_\sT^{\mu\nu}\frac{\pT^2}{2M_p^2}\bigg)
	h_1^{\perp\,g} \bigg \} + \text{HT},
\end{multline}
with $p_{\sT}^2 = -\pT^2$ and $g^{\mu\nu}_{\sT} = g^{\mu\nu}
- P^{\mu}P^{\prime\nu}/P{\cdot}P^\prime - P^{\prime\mu}P^{\nu}/P{\cdot}P^\prime$,
where $P$ and $P^\prime$ are the momenta of the colliding protons and $M_p$ their mass.
The gauge link $\mc{U}_{[0,\xi]}^{n[\text{--}]}$ in the matrix element runs from $0$ to $\xi$ 
via minus infinity along the direction $n$, which is a time-like dimensionless four-vector with no transverse
components such that $\zeta^2 = (2n{\cdot}P)^2/n^2$.
%the $n = P/\zeta + \zeta P^\prime/(2 P\cdot P^\prime)$ 
In principle, Eqs.\ \eqref{eq:factformula} and \eqref{eq:TMDcorrelator} also contain
soft factors, but with the appropriate choice of $\zeta$ (of around 1.5 times the hadronic center of 
mass energy), one can neglect their contribution, at least up to next-to-leading order \cite{Ji:2005nu,Ma:2012hh}.
The renormalization scale should be chosen around the characteristic scale of the hard
interaction.
The last line of Eq.\ \eqref{eq:TMDcorrelator} contains the parameterization of the
TMD correlator in terms of the unpolarized gluon distribution $f_1^g(x,\pT^2,\zeta,\mu)$,
the linearly polarized gluon distribution $h_1^{\perp\,g}(x,\pT^2,\zeta,\mu)$ and
Higher Twist (HT) terms, which only give $\mc{O}(1/Q)$ suppressed contributions to the
cross section, where $Q\equiv \sqrt{q^2}$.

The general structure of the differential cross section for the process $pp\to \gamma\gamma X$
is given by \cite{Qiu:2011ai}
\begin{multline}\label{eq:genstruc}
\frac{\d\sigma}{\d^4 q \d \Omega} 
  \propto 
	F_1(Q,\theta)\, \mc{C} \left[f_1^gf_1^g\right]
	+ F_2(Q,\theta)\,	\mc{C} \left[w_2\, h_1^{\perp g}h_1^{\perp g}\right]\\
	+ F_3(Q,\theta)\, 	\mc{C} \left[w_3 f_1^g h_1^{\perp g}
				  + (x_1 \leftrightarrow x_2) \right]\cos(2\phi)\\
	+ F_3^{\prime}(Q,\theta)\, \mc{C} \left[w_3 f_1^g h_1^{\perp g} 
				  - (x_1 \leftrightarrow x_2) \right]\sin(2\phi)\\
	+ F_4(Q,\theta)\,	\mc{C} \left[w_4\, h_1^{\perp g}h_1^{\perp g}\right]\cos(4\phi)
	+\mc{O}\left(\frac{q_\sT}{Q}\right),
\end{multline}
where the $F_i$ factors consist of specific combinations of $gg\to X_{0,2}\to \gamma\gamma$
helicity amplitudes, with $F_{3,4}$ involving amplitudes with opposite gluon helicities.
The convolution $\mc{C}$ is defined as
\begin{multline}
\mathcal{C}[w\, f\, g] \equiv \int\! \d^{2}\pT\int\! \d^{2}\kT\,
\delta^{2}(\pT+\kT-\bm q_{\sT})\\
w(\pT,\kT)\, f(x_{1},\pT^{2})\, g(x_{2},\kT^{2})
\end{multline}
and the weights appearing in the convolutions as
\begin{align}
 w_2		&\equiv \frac{2 (\kT{\cdot}\pT)^2 - \kT^2 \pT^2 }{4 M_p^4},\nonumber\\
 w_3 		&\equiv \frac{\qT^2\kT^2 - 2 (\qT{\cdot}\kT)^2}{2 M_p^2 \qT^2},\nonumber\\
 w_4		&\equiv  2\left[\frac{\pT{\cdot}\kT}{2M_p^2} - 
		\frac{(\pT{\cdot}\qT) (\kT{\cdot}\qT)}{M_p^2\qT^2}\right]^2 -\frac{\pT^2\kT^2 }{4 M_p^4}.
\end{align}

The TMD distribution functions contain both perturbative and non-perturbative information.
The tails ($\pT \gg M_p$) of the distribution functions can be calculated
using pQCD, but the low $\pT$ region will inevitably contain non-perturbative hadronic information.
To get a description over the full $\pT$ range one needs to extract the TMD
distribution functions from experimental data \cite{Qiu:2011ai,Boer:2010zf}.

\begin{figure}[htb]
\centering
\includegraphics[width=8.6cm]{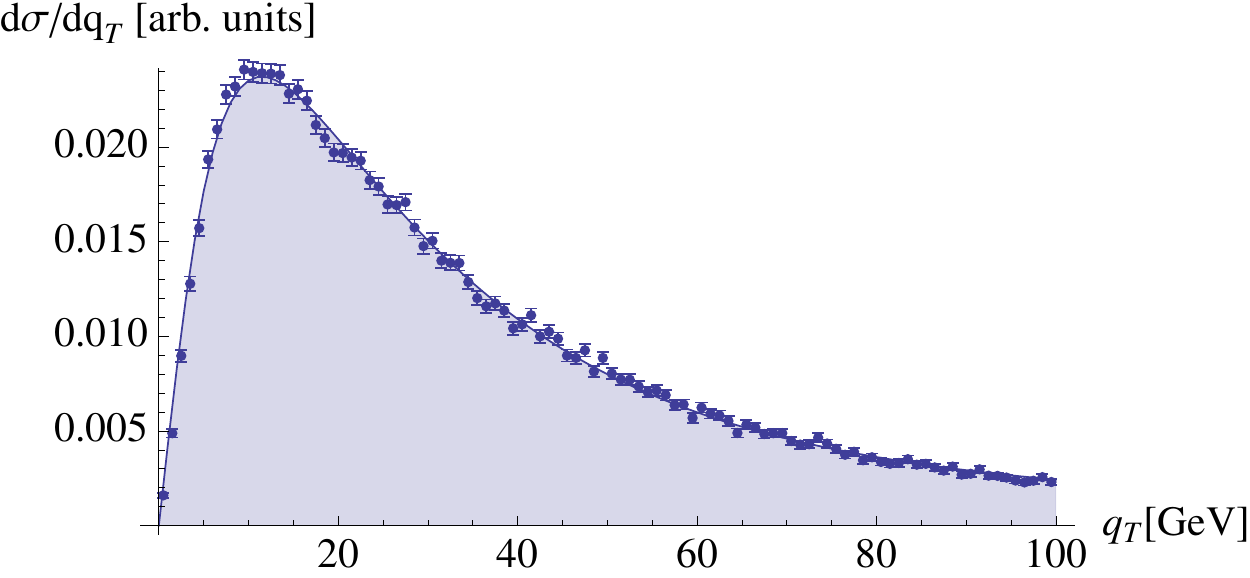} 
\caption{Plot of $q_\sT \mc{C}[f_1^g f_1^g]$ (line) and the Pythia Higgs 
$\d\sigma/\d q_\sT$ distribution for $M_h=125$ GeV at $\sqrt{s}=8$ TeV (points).}
\label{fig:qTCff}
\end{figure}

To make numerical predictions we will use a functional form for the unpolarized 
gluon TMD which has, in accordance with the pQCD calculation, a $1/\pT^2$ tail at large $\pT$
and resembles a Gaussian for small $\pT$,
\begin{equation}\label{eq:f1gparam}
 f_1^g(x,\pT^2,\frac{3}{2} \sqrt{s},M_h) = 
 \frac{ A_0^{}\, M_0^2}{M_0^2 + \pT^2} \exp\left[-\frac{\pT^2}{a \pT^2+2 \sigma ^2} \right]\!.
\end{equation}
Preferably one would fit the parameters in Eq.\ \eqref{eq:f1gparam} to actual data,
but since those are currently not available we will instead fit to the Standard Model Higgs boson 
transverse momentum distribution 
obtained by interfacing the POWHEG \cite{Nason:2004rx,Frixione:2007vw,Alioli:2008tz}
NLO gluon fusion calculation \cite{Alioli:2010xd} to Pythia 8.170 \cite{Sjostrand:2006za,Sjostrand:2007gs},
assuming a Higgs mass of 125 GeV and a collider center of mass energy of 8 TeV.
Pythia does not take into account effects of gluon polarization,
so we fit the data by setting the linearly polarized gluon distribution equal to zero.
In this way the TMD prediction without gluon polarization agrees with the Pythia prediction.
We think this is the most realistic choice we can make, 
because Pythia is tuned to reproduce collider data well.
Our Gaussian-with-tail Ansatz is able to adequately fit the Pythia data,
as is shown in Figure \ref{fig:qTCff}.
The fit results in the following values for the parameters $\sigma= 38.9$ GeV,
$a= 0.555$ and $M_0 = 3.90$ GeV.
We are not concerned about the overall normalization, as we will be
only interested in distributions and not the absolute size of the cross section.

The linearly polarized gluon distribution will be expressed in terms of the 
unpolarized gluon distribution and the degree of polarization $\mc{P}$, i.e.,
\begin{equation}
 h_1^{\perp g}(x,\pT,\zeta,\mu) = \mc{P}(x,\pT^2,\zeta) \frac{2M_p^2}{\pT^2} f_1^g(x,\pT,\zeta,\mu),
\end{equation}
such that $|\mc{P}|=1$ corresponds to $h_1^{\perp g}$ saturating its upper bound \cite{Mulders:2000sh}
and with the correct power law tail as first calculated in \cite{Sun:2011iw}.
Calculations of the gluon TMD distributions using the Color Glass Condensate model
predict maximal gluon polarization for large $\pT$ and small $x$ \cite{Metz:2011wb}.
Ideally one extracts the degree of polarization from data, but this is currently unfeasible. 

Perturbative QCD can be used to calculate the large $\pT$ tails of the TMD distributions 
in terms of the collinear parton distribution functions
as has been done in Ref. \cite{Ji:2005nu} for the unpolarized distribution and
Ref. \cite{Sun:2011iw} for the linearly polarized gluon distribution.
We will follow a similar approach, 
but keep finite $\zeta$ instead of taking the $\zeta\to\infty$ limit
and calculate the degree of polarization to leading order in  $\alpha_s$ 
from the MSTW 2008 collinear parton distributions \cite{Martin:2009iq} 
evaluated at a scale of $\mu=2$ GeV.

\begin{figure}[htb]
\centering
\includegraphics[width=8.6cm]{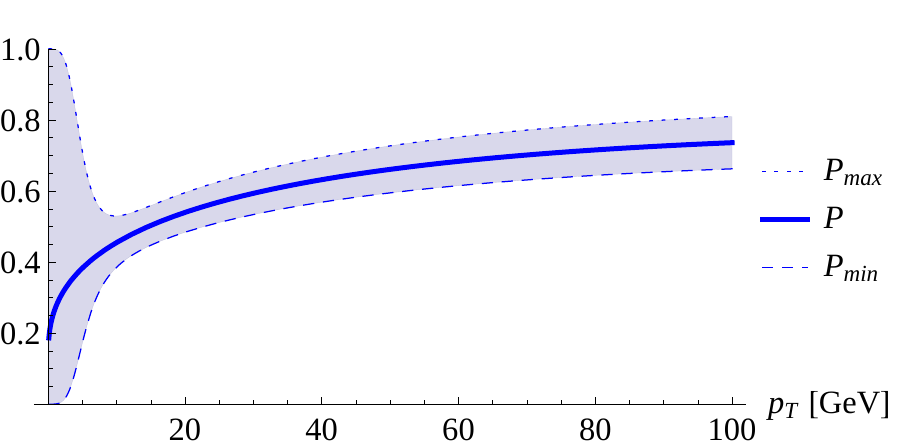} 
\caption{Plot of the degrees of polarization $\mc{P}_{min}$, $\mc{P}$ and $\mc{P}_{max}$
at $x=M_h/\sqrt{s}$, with $M_h=125$ GeV and $\sqrt{s}=8$ TeV.} 
\label{fig:P1P2P3}
\end{figure}

The pQCD calculation is only valid in the limit $\pT \gg M_p$. 
To model the lack of knowledge at low $\pT$,
we will define three different degrees of polarization 
$\mc{P}_{min}$, $\mc{P}$ and $\mc{P}_{max}$,
of which the first approaches zero at low $\pT$, the second follows the pQCD prediction
and the last reaches up to one at low $\pT$.
Other sources of uncertainty are the choices of the scales $\zeta$ and $\mu$ and the
omission of higher order terms.
We estimate this additional uncertainty, by varying the different scales,
to be maximally 10\% and model it by letting $\mc{P}_{max,min}$ approach the
pQCD calculation $\pm10\%$ for large $\pT$.
More specifically, we define
\begin{align}
 \mc{P}_{min}	&\equiv 
    \frac{\pT^4}{p_{0}^4 + \pT^4}\, 0.9\, \mc{P}_\text{pQCD}(x,\pT^2),\nonumber\\
 \mc{P} 	&\equiv 
    \mc{P}_\text{pQCD}(x,\pT^2),\nonumber\\
 \mc{P}_{max}	&\equiv 
    1- \frac{\pT^4}{p_{0}^4 + \pT^4}\left[1- 1.1\, \mc{P}_\text{pQCD}(x,\pT^2)\right],
\end{align}
where $ \mc{P}_\text{pQCD}$ is the pQCD degree of polarization calculated at
$\zeta=1.5\sqrt{s}$ and we take $p_0=5$ GeV.
The resulting $\mc{P}_{min}$, $\mc{P}$ and $\mc{P}_{max}$ are plotted in Figure \ref{fig:P1P2P3}.

We will consider the partonic process $gg\to X_{0,2}\to \gamma\gamma$ where $X$ is either
a spin-0 or spin-2 boson, with completely general couplings.
For the interaction vertex we will follow the conventions of Refs.\ \cite{Gao:2010qx} 
and \cite{Bolognesi:2012mm},
where the vertex coupling a spin-0 boson to massless gauge bosons is parameterized as
\begin{equation}
 V[X_0\to V^\mu (q_1) V^\nu (q_2)] = a_1 q^2 g^{\mu\nu} + a_3 \epsilon^{q_1 q_2 \mu\nu},
\end{equation}
and for a spin-2 boson as 
\begin{multline}
 V[X_2^{\alpha\beta}\to V^\mu (q_1)V^\nu (q_2)] = 
  \frac{1}{2}c_1 q^2 g^{\mu\alpha} g^{\nu\beta} \\
  + \left(c_2 q^2 g^{\mu\nu} + c_5 \epsilon^{q_1 q_2 \mu\nu} \right) \frac{\tilde{q}^\alpha \tilde{q}^\beta}{q^2},
\end{multline}
where $q\equiv q_1 + q_2$ and $\tilde{q}\equiv q_1 - q_2$.
The coupling to gluons can be different from the coupling to photons, but to keep expressions
compact we will consider them equal.

For the $gg\to X_0\to \gamma\gamma$ subprocess, the non-zero $F$ factors in Eq.\ \eqref{eq:genstruc} read
\begin{align}
 F_1 &= 16 |a_1|^4 + 8 |a_1|^2|a_3|^2 + |a_3|^2, \nonumber\\
 F_2 &= 16 |a_1|^4 - |a_3|^4,
\end{align}
and for the $gg\to X_2\to \gamma\gamma$ process one has
\begin{align}
 F_1 &= 18 A^+ |c_1|^2 s_{\theta}^4 +  {A^+}^2\! \left(1-3 c_{\theta}^2\right)^2	\nonumber\\
  &\phantom{=}+\frac{9}{8} |c_1|^4 (28 c_{2\theta}+c_{4\theta}+35),\nonumber\\
 F_2 &=  9 A^- |c_1|^2 s_{\theta}^4 +  A^-A^+\left(1-3 c_{\theta}^2\right)^2,\nonumber\\
 F_3 &= 3 s_{\theta}^2 B^- \left[3 |c_1|^2 (c_{2\theta}+3) + A^+(3 c_{2\theta}+1) \right],\nonumber\\
 F_3^\prime &= 6 s_{\theta}^2 \Re(c_1 c_5^*) 
    \left[3 |c_1|^2 (c_{2\theta}+3) + A^+(3 c_{2\theta}+1)\right],\nonumber\\
 F_4 &= 9  s_{\theta}^4 |c_1|^2 \left[2B^+ + 4 |c_5|^2\right],
\end{align}
where we have defined $A^\pm\equiv |c_1+4c_2|^2\pm 4|c_5|^2$, $B^\pm\equiv |c_1 + 2 c_2|^2\pm 4|c_2|^2$,
$c_{n\theta} \equiv \cos(n\theta)$ and $s_{\theta} \equiv \sin(\theta)$.
Overall factors have been dropped, because as said we will be
only interested in distributions and not the absolute size of the cross section.
Unlike the case for Higgs production from linearly polarized photons \cite{Grzadkowski:1992sa},
there is no direct observable signalling $CP$ violation in the spin-0 case.
For the spin-2 case there \emph{is} such a clear signature, 
being a $\sin2\phi$ dependence of the cross section, which can only be present if both
$c_1$ and $c_5$ are non-zero, implying a $CP$-violating interaction.

\begin{table}[htb]
\centering
 \begin{tabular}{cccccccc}
 scenario 	&$0^+$ &$0^-$ &$2_m^+$ &$2_h^+$ &$2_{h^\prime}^+$ &$2_{h^{\prime\prime}}^+$ &$2_{h}^-$\\
 \hline
 \hline
 $a_1$		&1 &0 &- &- &- &- &-   		\\
 $a_3$		&0 &1 &- &- &- &- &- 		\\
 $c_1$		&- &- &1 &0 &1 &1 &0  	\\
 $c_2$		&- &- &$-\frac{1}{4}$ &1 &1 &$-\frac{3}{2}$ &0  		\\
 $c_5$ 		&- &- &0 &0 &0 &0 &1  		\\
 \end{tabular} 
\caption{Different spin, parity and coupling scenarios.}\label{tab:bmscenarios}
\end{table} 

In Ref.\ \cite{Bolognesi:2012mm} a set of different spin, parity and coupling scenarios 
is defined.
To those scenarios we will add $2_{h^\prime}^+$ and $2_{h^{\prime\prime}}^+$, 
which will serve as examples of higher-dimensional spin-2 coupling hypotheses that are 
indistinguishable in the $\theta$ distribution, but do have a different $\phi$ distribution.
The scenarios are summarized in Table \ref{tab:bmscenarios}.

\begin{figure}[htb]
\centering
\includegraphics[width=8.6cm]{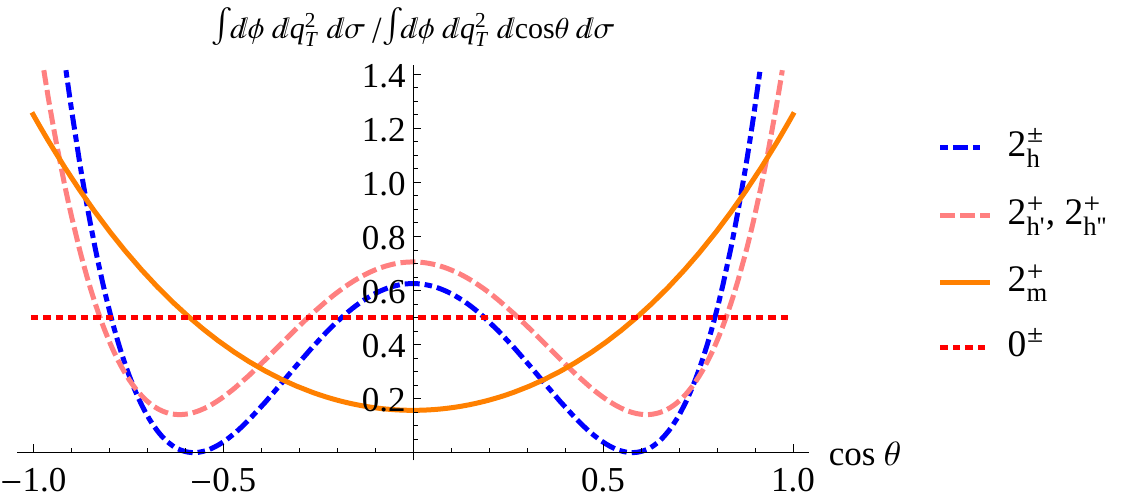}
\caption{Plot of the $\cos\theta$ distribution for the various scenarios.}
\label{fig:dcosthetadistr}
\end{figure}

In Figure \ref{fig:dcosthetadistr} we show the diphoton $\cos\theta$ distribution
 for the various scenarios.
Looking only at this distribution $0^+$ and $0^-$ are indistinguishable, 
as are $2_h^+$ and $2_h^-$, and also $2_{h^\prime}^+$ and $2_{h^{\prime\prime}}^+$. 

\begin{figure}[htb]
\centering
\includegraphics[width=8.6cm]{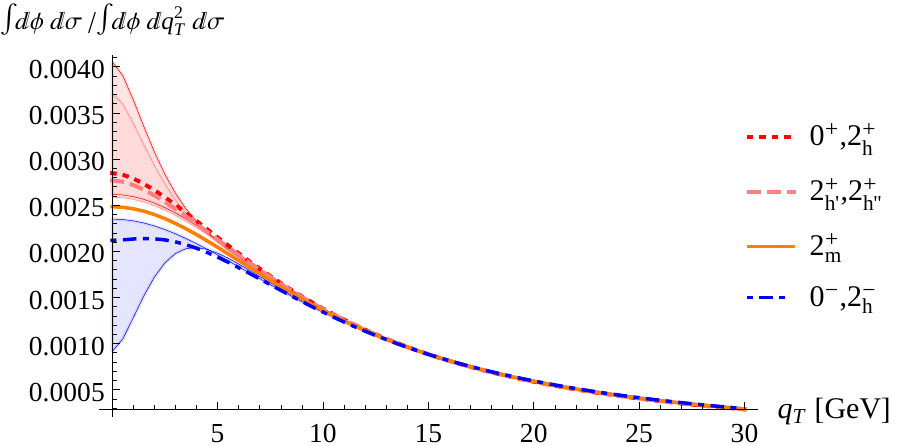}
\caption{Plot of the $q_\sT$ distribution
for the various coupling schemes at $\theta=\pi/2$ and zero rapidity, 
using an upper limit on the $q_\sT$ integration in the denominator of $M_h/2$.
The shaded area is due to the uncertainty in the degree of polarization.}
\label{fig:qTsqdistr}
\end{figure}

In Figure \ref{fig:qTsqdistr} we show the diphoton transverse momentum distribution
for the different coupling hypotheses at fixed $\theta=\pi/2$ and at zero rapidity.
The positive parity states show an enhancement at low $q_\sT$ ($<15$ GeV)
with respect to the negative parity states. 
At high $q_\sT$ ($>15$ GeV) this is reversed,
but with such a strongly reduced magnitude that it is invisible in the plot.
The $q_\sT$ distribution can thus, in principle, be used to determine
the parity of the newly found boson \cite{Boer:2011kf,denDunnen:2012ym}.
Although the difference is small and most likely difficult to measure experimentally,
 this is the only way we know to determine the parity in the 
$gg \to X_{0,2} \to \gamma\gamma$ channel.

\begin{figure}[htb]
\centering
\includegraphics[width=8.6cm]{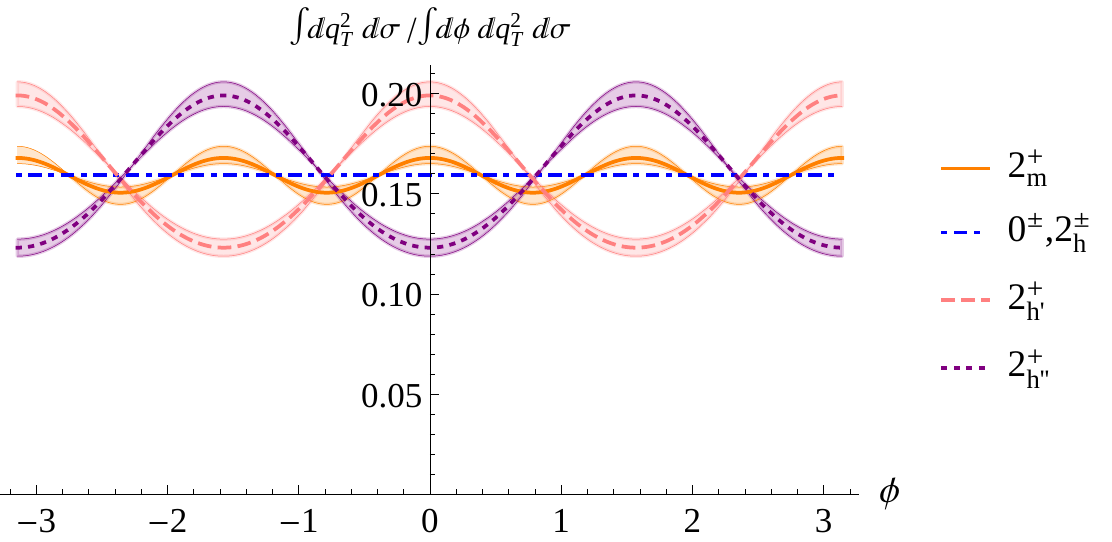}
\caption{Plot of the $\phi$ distribution 
for the different benchmark scenarios at $\theta=\pi/2$ and zero rapidity, 
using an upper limit on the $q_\sT$ integration of $M_h/2$.
The shaded area is due to the uncertainty in the degree of polarization.}
\label{fig:phidistr}
\end{figure}

Figure \ref{fig:phidistr} shows the diphoton $\phi$ distribution for the selected scenarios
at fixed $\theta=\pi/2$ and at zero rapidity.
The scalar, pseudoscalar and $2_h^\pm$ hypotheses show a uniform $\phi$ distribution, 
whereas the $2_m^+$ has a characteristic $\cos(4\phi)$ dependence
with an amplitude of $5.4_{-1.8}^{+3.7}\%$.
The $2_{h^\prime}^+$ and $2_{h^{\prime\prime}}^+$ scenarios
exhibit a weak $\cos(4\phi)$ modulation with an amplitude of $1.2_{-0.4}^{+0.8}\%$
and a strong $\cos(2\phi)$ modulation with an amplitude of $24\pm 3\%$ and opposite
sign.
The $\phi$ distribution thus offers a way to distinguish $0^\pm$, $2_m^+$, $2_{h^\prime}^+$ and 
$2_{h^{\prime\prime}}^+$ from each other,
something that is impossible with the $\cos\theta$ distribution alone.

We want to stress again that a $\sin2\phi$ dependence implies a $CP$-violation coupling,
which is thus very interesting to search for.
Note however that Higgs bosons produced with positive and negative rapidity have to be treated separately,
because those regions will have an opposite sign $\sin2\phi$ modulation and would otherwise cancel.
We also want to mention that $gg\to\gamma\gamma$ continuum production has a non-isotropic $\phi$ dependence,
with an amplitude approximately a factor 3 smaller than resonance production \cite{Qiu:2011ai,Thesis},
which should not be mistaken for a spin-2 Higgs.

In conclusion, we have calculated the diphoton distribution in the decay of arbitrary
spin-0 and spin-2 bosons produced from gluon fusion, taking into account the fact that 
gluons inside an unpolarized proton are generally linearly polarized.
The gluon polarization brings about a difference in the transverse momentum distribution 
of positive and negative parity states.
At the same time, it causes the azimuthal CS angle $\phi$ distribution
to be non-isotropic for various spin-2 coupling hypotheses.
These distributions allow spin and parity scenarios to be distinguished 
that cannot be done with the polar angle $\theta$ dependence alone.
We think that these observables could therefore form a valuable addition to the analysis methods 
to determine the spin, parity and coupling of the newly found boson at the LHC.

%%%%%%%%%%%%%%%%%%%%%%%%%%%%%%%%%%%%%%%%%%%%%%%%%%%%%%%%%%%%%%%%%%%%%%%
\begin{acknowledgments}
This work was supported in part by the German Bundesministerium f\"{u}r Bildung und Forschung (BMBF),
grant no. 05P12VTCTG.
\end{acknowledgments}
%%%%%%%%%%%%%%%%%%%%%%%%%%%%%%%%%%%%%%%%%%%%%%%%%%%%%%%%%%%%%%%%%%


\begin{thebibliography}{99}

%\cite{:2012gk}
\bibitem{:2012gk} 
  G.~Aad {\it et al.}  [ATLAS Collaboration],
  %``Observation of a new particle in the search for the Standard Model Higgs boson with the ATLAS detector at the LHC,''
  Phys.\ Lett.\ B {\bf 716}, 1 (2012)
  [arXiv:1207.7214 [hep-ex]].
  %%CITATION = ARXIV:1207.7214;%%
  
%\cite{:2012gu}
\bibitem{:2012gu} 
  S.~Chatrchyan {\it et al.}  [CMS Collaboration],
  %``Observation of a new boson at a mass of 125 GeV with the CMS experiment at the LHC,''
  Phys.\ Lett.\ B {\bf 716}, 30 (2012)
  [arXiv:1207.7235 [hep-ex]].
  %%CITATION = ARXIV:1207.7235;%%

\bibitem{ATLAS-gammagammaupdate}
\url{http://cds.cern.ch/record/1527124/files/ATLAS-CONF-2013-029.pdf}
  
\bibitem{ATLAS-ZZupdate}
\url{http://cds.cern.ch/record/1523699/files/ATLAS-CONF-2013-013.pdf}

\bibitem{ATLAS-WWupdate}
\url{http://cds.cern.ch/record/1527127/files/ATLAS-CONF-2013-031.pdf}

\bibitem{CMS-ZZupdate}
\url{http://cds.cern.ch/record/1523767/files/HIG-13-002-pas.pdf}

\bibitem{CMS-WWupdate}
\url{http://cds.cern.ch/record/1523673/files/HIG-13-003-pas.pdf}


%\cite{Landau:1948kw}
\bibitem{Landau:1948kw} 
  L.~D.~Landau,
  %``On the angular momentum of a two-photon system,''
  Dokl.\ Akad.\ Nauk Ser.\ Fiz.\  {\bf 60}, 207 (1948).
  %%CITATION = DANKA,60,207;%%
  %156 citations counted in INSPIRE as of 18 Mar 2013
  
%\cite{Yang:1950rg}
\bibitem{Yang:1950rg} 
  C.~-N.~Yang,
  %``Selection Rules for the Dematerialization of a Particle Into Two Photons,''
  Phys.\ Rev.\  {\bf 77}, 242 (1950).
  %%CITATION = PHRVA,77,242;%%
  %572 citations counted in INSPIRE as of 18 Mar 2013


%\cite{Ellis:2012jv}
\bibitem{Ellis:2012jv} 
  J.~Ellis, R.~Fok, D.~S.~Hwang, V.~Sanz and T.~You,
  %``Distinguishing `Higgs' Spin Hypotheses using gamma gamma and WW* Decays,''
  arXiv:1210.5229 [hep-ph].
  %%CITATION = ARXIV:1210.5229;%%
  %10 citations counted in INSPIRE as of 18 Mar 2013
  
%\cite{Gao:2010qx}
\bibitem{Gao:2010qx} 
  Y.~Gao, A.~V.~Gritsan, Z.~Guo, K.~Melnikov, M.~Schulze and N.~V.~Tran,
  %``Spin determination of single-produced resonances at hadron colliders,''
  Phys.\ Rev.\ D {\bf 81}, 075022 (2010)
  [arXiv:1001.3396 [hep-ph]].
  %%CITATION = ARXIV:1001.3396;%%
  
%\cite{Bolognesi:2012mm}
\bibitem{Bolognesi:2012mm} 
  S.~Bolognesi, Y.~Gao, A.~V.~Gritsan, K.~Melnikov, M.~Schulze, N.~V.~Tran and A.~Whitbeck,
  %``On the spin and parity of a single-produced resonance at the LHC,''
  Phys.\ Rev.\ D {\bf 86}, 095031 (2012)
  [arXiv:1208.4018 [hep-ph]].
  %%CITATION = ARXIV:1208.4018;%%

%\cite{Collins:1977iv}
\bibitem{Collins:1977iv} 
  J.~C.~Collins and D.~E.~Soper,
  %``Angular Distribution of Dileptons in High-Energy Hadron Collisions,''
  Phys.\ Rev.\ D {\bf 16}, 2219 (1977).
  %%CITATION = PHRVA,D16,2219;%%
  %482 citations counted in INSPIRE as of 05 Mar 2013
  
  
%\cite{Boer:2011kf}
\bibitem{Boer:2011kf} 
  D.~Boer, W.~J.~den Dunnen, C.~Pisano, M.~Schlegel and W.~Vogelsang,
  %``Linearly Polarized Gluons and the Higgs Transverse Momentum Distribution,''
  Phys.\ Rev.\ Lett.\  {\bf 108}, 032002 (2012)
  [arXiv:1109.1444 [hep-ph]].
  %%CITATION = ARXIV:1109.1444;%%
  %14 citations counted in INSPIRE as of 19 Feb 2013
  
%\cite{denDunnen:2012ym}
\bibitem{denDunnen:2012ym} 
  W.~J.~den Dunnen, D.~Boer, C.~Pisano, M.~Schlegel and W.~Vogelsang,
  %``Linearly polarized Gluons and the Higgs Transverse Momentum Distribution,''
  arXiv:1205.6931 [hep-ph].
  %%CITATION = ARXIV:1205.6931;%%
  %2 citations counted in INSPIRE as of 19 Feb 2013

%\cite{Hagiwara:2009wt}
\bibitem{Hagiwara:2009wt} 
  K.~Hagiwara, Q.~Li and K.~Mawatari,
  %``Jet angular correlation in vector-boson fusion processes at hadron colliders,''
  JHEP {\bf 0907}, 101 (2009)
  [arXiv:0905.4314 [hep-ph]].
  %%CITATION = ARXIV:0905.4314;%%
  %35 citations counted in INSPIRE as of 26 Mar 2013

%\cite{Andersen:2010zx}
\bibitem{Andersen:2010zx} 
  J.~R.~Andersen, K.~Arnold and D.~Zeppenfeld,
  %``Azimuthal Angle Correlations for Higgs Boson plus Multi-Jet Events,''
  JHEP {\bf 1006}, 091 (2010)
  [arXiv:1001.3822 [hep-ph]].
  %%CITATION = ARXIV:1001.3822;%%
  %12 citations counted in INSPIRE as of 26 Mar 2013

 %\cite{Frank:2012wh}
\bibitem{Frank:2012wh} 
  J.~Frank, M.~Rauch and D.~Zeppenfeld,
  %``Spin-2 Resonances in Vector-Boson-Fusion Processes at NLO QCD,''
  arXiv:1211.3658 [hep-ph].
  %%CITATION = ARXIV:1211.3658;%%
  %7 citations counted in INSPIRE as of 18 Mar 2013

  
%\cite{Sun:2011iw}
\bibitem{Sun:2011iw} 
  P.~Sun, B.~-W.~Xiao and F.~Yuan,
  %``Gluon Distribution Functions and Higgs Boson Production at Moderate Transverse Momentum,''
  Phys.\ Rev.\ D {\bf 84}, 094005 (2011)
  [arXiv:1109.1354 [hep-ph]].
  %%CITATION = ARXIV:1109.1354;%%
  %12 citations counted in INSPIRE as of 21 Feb 2013

%\cite{Ma:2012hh}
\bibitem{Ma:2012hh} 
  J.~P.~Ma, J.~X.~Wang and S.~Zhao,
  %``TMD Factorization for Quarkonium Production at Low Transverse Momentum,''
  arXiv:1211.7144 [hep-ph].
  %%CITATION = ARXIV:1211.7144;%%
  %1 citations counted in INSPIRE as of 21 Feb 2013

%\cite{Ji:2005nu}
\bibitem{Ji:2005nu} 
  X.~-d.~Ji, J.~-P.~Ma and F.~Yuan,
  %``Transverse-momentum-dependent gluon distributions and semi-inclusive processes at hadron colliders,''
  JHEP {\bf 0507}, 020 (2005)
  [hep-ph/0503015].
  %%CITATION = HEP-PH/0503015;%%
  %48 citations counted in INSPIRE as of 21 Feb 2013
  
  
%\cite{Qiu:2011ai}
\bibitem{Qiu:2011ai} 
  J.~-W.~Qiu, M.~Schlegel and W.~Vogelsang,
  %``Probing Gluonic Spin-Orbit Correlations in Photon Pair Production,''
  Phys.\ Rev.\ Lett.\  {\bf 107}, 062001 (2011)
  [arXiv:1103.3861 [hep-ph]].
  %%CITATION = ARXIV:1103.3861;%%
  %18 citations counted in INSPIRE as of 05 Mar 2013
  
%\cite{Boer:2010zf}
\bibitem{Boer:2010zf} 
  D.~Boer, S.~J.~Brodsky, P.~J.~Mulders and C.~Pisano,
  %``Direct Probes of Linearly Polarized Gluons inside Unpolarized Hadrons,''
  Phys.\ Rev.\ Lett.\  {\bf 106}, 132001 (2011)
  [arXiv:1011.4225 [hep-ph]].
  %%CITATION = ARXIV:1011.4225;%%
  %23 citations counted in INSPIRE as of 25 Mar 2013
  
%\cite{Alioli:2008tz}
\bibitem{Alioli:2008tz} 
  S.~Alioli, P.~Nason, C.~Oleari and E.~Re,
  %``NLO Higgs boson production via gluon fusion matched with shower in POWHEG,''
  JHEP {\bf 0904}, 002 (2009)
  [arXiv:0812.0578 [hep-ph]].
  %%CITATION = ARXIV:0812.0578;%%
  %107 citations counted in INSPIRE as of 18 Mar 2013
  
%\cite{Nason:2004rx}
\bibitem{Nason:2004rx} 
  P.~Nason,
  %``A New method for combining NLO QCD with shower Monte Carlo algorithms,''
  JHEP {\bf 0411}, 040 (2004)
  [hep-ph/0409146].
  %%CITATION = HEP-PH/0409146;%%
  %384 citations counted in INSPIRE as of 18 Mar 2013

%\cite{Frixione:2007vw}
\bibitem{Frixione:2007vw} 
  S.~Frixione, P.~Nason and C.~Oleari,
  %``Matching NLO QCD computations with Parton Shower simulations: the POWHEG method,''
  JHEP {\bf 0711}, 070 (2007)
  [arXiv:0709.2092 [hep-ph]].
  %%CITATION = ARXIV:0709.2092;%%
  %491 citations counted in INSPIRE as of 18 Mar 2013

%\cite{Alioli:2010xd}
\bibitem{Alioli:2010xd} 
  S.~Alioli, P.~Nason, C.~Oleari and E.~Re,
  %``A general framework for implementing NLO calculations in shower Monte Carlo programs: the POWHEG BOX,''
  JHEP {\bf 1006}, 043 (2010)
  [arXiv:1002.2581 [hep-ph]].
  %%CITATION = ARXIV:1002.2581;%%
  %222 citations counted in INSPIRE as of 18 Mar 2013

 
%\cite{Sjostrand:2006za}
\bibitem{Sjostrand:2006za} 
  T.~Sjostrand, S.~Mrenna and P.~Z.~Skands,
  %``PYTHIA 6.4 Physics and Manual,''
  JHEP {\bf 0605}, 026 (2006)
  [hep-ph/0603175].
  %%CITATION = HEP-PH/0603175;%%
  
%\cite{Sjostrand:2007gs}
\bibitem{Sjostrand:2007gs} 
  T.~Sjostrand, S.~Mrenna and P.~Z.~Skands,
  %``A Brief Introduction to PYTHIA 8.1,''
  Comput.\ Phys.\ Commun.\  {\bf 178}, 852 (2008)
  [arXiv:0710.3820 [hep-ph]].
  %%CITATION = ARXIV:0710.3820;%%

  
%\cite{Mulders:2000sh}
\bibitem{Mulders:2000sh} 
  P.~J.~Mulders and J.~Rodrigues,
  %``Transverse momentum dependence in gluon distribution and fragmentation functions,''
  Phys.\ Rev.\ D {\bf 63}, 094021 (2001)
  [hep-ph/0009343].
  %%CITATION = HEP-PH/0009343;%%
  %55 citations counted in INSPIRE as of 20 Mar 2013
  
%\cite{Metz:2011wb}
\bibitem{Metz:2011wb} 
  A.~Metz and J.~Zhou,
  %``Distribution of linearly polarized gluons inside a large nucleus,''
  Phys.\ Rev.\ D {\bf 84}, 051503 (2011)
  [arXiv:1105.1991 [hep-ph]].
  %%CITATION = ARXIV:1105.1991;%%
  
%\cite{Martin:2009iq}
\bibitem{Martin:2009iq} 
  A.~D.~Martin, W.~J.~Stirling, R.~S.~Thorne and G.~Watt,
  %``Parton distributions for the LHC,''
  Eur.\ Phys.\ J.\ C {\bf 63}, 189 (2009)
  [arXiv:0901.0002 [hep-ph]].
  %%CITATION = ARXIV:0901.0002;%%
  %1382 citations counted in INSPIRE as of 02 Apr 2013

%\cite{Grzadkowski:1992sa}
\bibitem{Grzadkowski:1992sa} 
  B.~Grzadkowski and J.~F.~Gunion,
  %``Using back scattered laser beams to detect CP violation in the neutral Higgs sector,''
  Phys.\ Lett.\ B {\bf 294}, 361 (1992)
  [hep-ph/9206262].
  %%CITATION = HEP-PH/9206262;%%
  %133 citations counted in INSPIRE as of 19 Mar 2013
  
\bibitem{Thesis}
  W.~J.~den Dunnen,
  ``Polarization effects in proton-proton collisions within the Standard Model and beyond,''
  (PhD Thesis) (2012)
  \url{http://dare.ubvu.vu.nl/handle/1871/39659}
  
\end{thebibliography}
\end{document}